\documentstyle[12pt]{article}
\title{Construction of the Space of States for
       Axiomatic Scattering Theory}
\author{Oskar Pelc \and L. P. Horwitz}

\newcommand{\newsection}[1]{
 \vspace{10mm} \pagebreak[3]
 \addtocounter{section}{1}
 \setcounter{equation}{0}
 \setcounter{subsection}{0}
 \setcounter{paragraph}{0}
 \setcounter{equation}{0}
 \setcounter{figure}{0}
 \setcounter{table}{0}
 \addcontentsline{toc}{section}{\protect\numberline{\arabic{section}}{#1}}
 \begin{flushleft}
  {\large\bf \thesection. #1}
 \end{flushleft}
 \nopagebreak}

\renewcommand{\thesection}{\Roman{section}}

\newcommand{\qed}{\hfill \rule{2.5mm}{2.5mm}}
\newcommand{\al}{\alpha}

\newcommand{\Ag}{A_{g}}
\newcommand{\AP}{{\cal A_{P}}}
\newcommand{\beq}{\begin{equation}}
\newcommand{\bbar}{\overline}

\newcommand{\dg}{^{\dagger}}
\newcommand{\der}[1]{\frac{d}{d#1}}
\newcommand{\derp}[1]{\frac{\partial}{\partial#1}}
\newcommand{\dl}{\delta}
\newcommand{\D}{{\cal D}}
\newcommand{\Dl}{\Delta}
\newcommand{\eeq}{\end{equation}}
\newcommand{\eg}{{\em e.g.\ }}
\newcommand{\E}{{\cal E}}
\newcommand{\f}{\varphi}

\newcommand{\F}{\Phi}
\newcommand{\Fa}{\F_{\al}}

\newcommand{\Fn}{\F^{(n)}}
\newcommand{\Fnd}{\F^{(n)'}}
\newcommand{\Fns}{\Fn_{s}}
\newcommand{\Fnsd}{\Fnd_{s}}
\newcommand{\Fone}{\F^{(1)}}
\newcommand{\Foned}{\F^{(1)'}}
\newcommand{\Fs}{\F_{s}}
\newcommand{\Ha}{\Hc_{\al}}
\newcommand{\Hc}{{\cal H}}
\newcommand{\Hn}{\Hc^{(n)}}
\newcommand{\Hns}{\Hn_{s}}
\newcommand{\Hone}{\Hc^{(1)}}
\newcommand{\Hs}{\Hc_{s}}
\newcommand{\ie}{{\em i.e.\ }}
\newcommand{\lm}{\lambda}
\newcommand{\inv}{^{-1}}
\newcommand{\La}{L^{\al}}
\newcommand{\Lc}{{\cal L}}
\newcommand{\Lm}{\Lambda}
\newcommand{\Lmp}{\Lm_{p}}
\newcommand{\Lp}{L^{(p)}}
\newcommand{\Lsq}{\Lc^{2}}
\newcommand{\m}{^{m}}
\newcommand{\ma}{m_{\al}}
\newcommand{\mh}{\hat{\mu}}
\newcommand{\mm}{\mu_{m}}
\newcommand{\mma}{\mu_{\ma}}
\newcommand{\mn}{^{(m,n)}}
\newcommand{\ms}{^{(m)}}
\newcommand{\M}{{\cal M}}
\newcommand{\n}{^{n}}
\newcommand{\na}{\nu_{\al}}
\newcommand{\ns}{^{(n)}}

\newcommand{\Om}{\Omega}
\newcommand{\pma}{p_{\ma}}

\newcommand{\po}{p_{0}}
\newcommand{\psm}{p_{m}}
\newcommand{\pv}{\vec{p}}
\newcommand{\Pa}{\Psi_{\al}}
\newcommand{\Pc}{{\cal P}}
\newcommand{\Pm}{P^{\mu}}
\newcommand{\r}{\rho}
\newcommand{\ra}{\r(\al)}
\newcommand{\s}{\sigma}
\newcommand{\sa}{\s_{\al}}
\newcommand{\so}{_{0}}

\newcommand{\T}{\hat{\cal T}}
\newcommand{\TF}{\T_{F}}
\newcommand{\Tm}{\T_{m}}
\newcommand{\Tma}{\T_{\ma}}
\newcommand{\Ua}{U^{\al}}
\newcommand{\Uone}{U^{(1)}}

\newcommand{\UP}{{\cal U_{P}}}
\newcommand{\Ut}{\tilde{U}}

\def\al{\alpha}

\def\dl{\delta}                \def\Dl{\Delta}

\def\lm{\lambda}               \def\Lm{\Lambda}

               \def\Om{\Omega}
               \def\Sg{\Sigma}

\def\Pc{\mbox{\protect$\cal P$}}

\def\inv{^{-1}}

\def\hs{\hspace{5mm}}
\def\hsc{\hspace{5mm},\hspace{5mm}}
\def\ie{{\em i.e.\ }}
\def\eg{{\em e.g.\ }}

\def\beq{\begin{equation}}
\def\eeq{\end{equation}}

\begin{document}


\begin{titlepage}

\begin{flushright}
RI-1-96 \\
TAUP 2312-95\\
IASSNS-96/40\\
\today
\\[15mm]
\end{flushright}

\begin{center}
\Large
Construction of a Complete Set of States \\
in Relativistic Scattering Theory
\\[10mm]
\large
Oskar Pelc$^*$\footnote{E-mail: oskar@shum.cc.huji.ac.il}
\normalsize and \large 
L. P. Horwitz$\dg$\footnote{
    E-mail: horwitz@sns.ias.edu

 On sabbatical leave from the School of Physics and Astronomy, 
 Raymond and Beverly Sackler Faculty of Exact Sciences,
 Tel-Aviv University, Ramat-Aviv, Israel;  also at Department of Physics, 
 Bar-Ilan University, Ramat Gan, Israel.}
\normalsize \\[5mm]
{\em $^*$Racah Institute of Physics, The Hebrew University\\
  Jerusalem, 91904, Israel}
\normalsize \\[5mm]
{\em $\dg$School of Natural Sciences, Institute for Advanced Study,\\
 Princeton, N.J. 08540, USA}
\\[15mm]
\end{center}

\begin{abstract}
The space of physical states in relativistic scattering theory is
constructed, using a rigorous version of the Dirac formalism, where the 
Hilbert space structure is extended to a Gel'fand triple. 
This extension enables the construction
of ``a complete set of states'', the basic concept of the original Dirac
formalism, also in the cases of unbounded operators and continuous spectra.
We construct explicitly the Gel'fand triple and a complete set of ``plane
waves'' -- momentum eigenstates -- using the group of space-time symmetries.
This construction is
used (in a separate article) to prove a generalization of the
Coleman-Mandula theorem to higher dimension.
\end{abstract}

------------------------------

PACS codes: 3.65.Db, 11.10.Cd, 11.30.Cp, 11.55.-m, 11.80.-m

\end{titlepage}

\pagestyle{plain}
\setcounter{page}{1}
\flushbottom

\newsection{Introduction}

Scattering experiments are among the main sources of experimental
information about the fundamental interactions in the sub-atomic range,
therefore, any theory that is intended to describe these interactions,
should provide predictions of scattering amplitudes. Much information
about these amplitudes can be obtained using only very basic and well
established assumptions as, for example, the fundamental postulates of
quantum mechanics. This is the approach of {\em the theory of the
S-matrix} (see, for example, ref. \cite{Iagolnitzer}). Results
obtained with such an approach naturally have a wide range of validity and
are applicable in any theory of fundamental interaction. They can be
derived from the present standard model, but are also expected to remain
valid when an improved fundamental description is found.

In the study of general properties of quantum mechanics it is natural to
use Dirac's ``bra'' and ``ket'' formalism \cite{Dirac}, because of its
remarkable elegance and simplicity. The essence of its usefulness is the
use of {\em a complete set of states} $\{<a|\}$ to form a representation
of the unity operator
\beq I\equiv\Sg_a|a><a| \eeq
which then can be used to represent various expression in terms of vector
components $<a|\psi>$ and matrix elements $<a'|A|a>$ of operators. In
scattering scenarios, a key role is played by the momentum operator and one
would like to ``diagonalize'' it, \ie, to take as a complete set of states,
a set of its eigenvectors (``plane waves''). However, the spectrum of the
momentum operator is continuous and the application of the Dirac formalism,
in its original form, to such operators is not well defined, since they do
not have a complete set of eigenvectors in the usual sense. This formalism
is, therefore, unsuitable for rigorous analysis of scattering theory.

An improved version of the Dirac formalism was developed \cite{Roberts-Bohm} 
\cite{Antoine1} \cite{Antoine2}, using the theory of distributions \cite{RHS}.
In this formalism, the Hilbert space structure is extended to a {\em Gel'fand 
triple} $(\Phi,\Hc,\Phi')$ (called also {\em  Rigged Hilbert Space}), where 
$\Hc$ is the Hilbert space of states, $\Phi$ is
a dense subspace of $\Hc$ and $\Phi'$ is the {\em dual} of $\Phi$ -- the
space of continuous linear functionals on $\Phi$. With an appropriate
choice of $\Phi$, a (generalized) complete set of eigenvectors of an
operator with a continuous spectrum can be found among elements of $\Phi'$
and most of the elegance of the original formalism can be recovered.

In this work we use the improved formalism (following the presentation of 
Antoine \cite{Antoine1} \cite{Antoine2}), to construct a complete set of
``plane waves'' for relativistic scattering theory. In the spirit of the
theory of the S-matrix, we use only very basic assumptions and state them
explicitly, to make apparent the range of validity of the results.
The construction relies on the symmetry of the theory under the group of
(restricted) space-time transformations: translations, rotations and boosts.
We consider an arbitrary dimension of space-time and assume the symmetry
group to be (isomorphic to) ${\Pc}(r,s)$ -- the inhomogeneous pseudo
orthogonal group of signature $(r,s)$ -- with arbitrary $r,s$. Eventually
we restrict to $s=1$ and the complete space of ``plane waves'' is
constructed for this case.

The structure of this article is as follows:
In Section 2, the space of states is realized as a space of functions over
momentum space; in Section 3 a Gel'fand triple and a ``complete set'' of
``plane waves'' are constructed. Section 4 illustrates the use of the
construction by rederiving some familiar relations and formulas 
(a further use of this construction is made in ref. \cite{Horwitz-Pelc}).
Finally, in Section 5, we comment on the assumptions made and on possible
extensions. Appendix A provides a concise description of the Dirac
formalism used in this work. Appendix B summarizes some relevant
properties of ${\cal P}(r,s)$.

\newsection{The Space of states}

In this Section we construct an explicit realization of the space of
states as a space of functions over momentum space, using the space-time
symmetry. In relativistic scattering theory, the space $\Hs$ of physical
states is a direct sum of (complex, separable) Hilbert spaces
\begin{equation}
  \Hs=\bigoplus_{n=0}^{\infty}\Hns
\end{equation}
where $\Hns$ is the space of $n$-particle states (thus
called ``$n$-particle space'') and is (isomorphic to) a closed subspace
of the completed tensor product of $n$ one-particle spaces:
\begin{equation}\label{Hns}
  \Hns\subset\Hn=\bbar{\bigotimes_{1}^{n}\Hone}.
\end{equation}
The elements of $\Hns$ are those elements of
$\Hn$ which have the right symmetry properties with respect
to the exchange of identical particles.

The S-matrix $S$ is assumed to be a unitary operator on $\Hs$.

\subsection{The Poincar\'{e} Symmetry in $\Hone$}
\label{Psymm}

A symmetry transformation of the S-matrix is defined to be a unitary
or antiunitary operator $U$ in $\Hs$ which satisfies:
\begin{enumerate}
  \item $\Hone$ is $U$-invariant, {\em i.e.\ }$U$ turns one-particle
    states into one-particle states;
  \item $U$ acts on many-particle states in accordance to their
    relation to the tensor product of one-particle states:
    \begin{equation}
      U(f_{1}\otimes\cdots\otimes f_{n})=
        (Uf_{1})\otimes\cdots\otimes(Uf_{n})
    \end{equation}
    (and thus, according to property 1, $\forall n,\;\Hns$ is
    $U$-invariant);
  \item $U$ commutes with $S$.
\end{enumerate}
The invariance under (restricted) space-time transformations implies:

\begin{quote}
  \bf Assumption 2.1:

  \em There exists a connected group $\Pc_0'$ of symmetries of $S$ which
  is locally isomorphic to $\Pc(r,s)$.
\end{quote}
The connectedness implies, among other things, that $\Pc_0'$ doesn't
contain antiunitary elements. According to property 1 of the symmetries of
$S$, $\Hone$ is $\Pc'\so$-invariant and thus constitutes a representation
space of $\Pc'\so$. Any representation of $\Pc'\so$ is naturally also a
representation of the universal covering group $\Pc$ of $\Pc'\so$
which is (because of the local isomorphism) globally isomorphic to the
universal covering group of the identity component (the largest
connected subgroup) $\Pc\so\equiv\Pc\so(r,s)$ of $\Pc(r,s)$
\mbox{\cite[p. 70]{Barut}} \cite{isomorph}.
Wigner \cite{Wigner} and Bargmann \cite{Bargmann} showed that
this representation $\Uone$ is in general a (strongly) continuous
unitary projective representation (called also ``ray representation''
or ``representation up to a phase'').
\begin{quote}
  \bf Assumptions: \em
  \begin{description}
    \item[2.2] $\Uone$ is a true representation of $\Pc$ (that is with no
      extra phase \cite{Barg-ray}.
    \item[2.3] $\Uone$ has only type I factors \cite[p. 145]{Barut} 
      \cite{As23}.
  \end{description}
The second assumption means that $\Uone$ is expressible in terms of
irreducible representations and since these are identified
with particle types (see the discussion in the introduction of
\cite{Horwitz-Pelc}), this requirement is actually part of the
physical interpretation..

\end{quote}
As explained in appendix B.1, the most general such
representation is of the following form (as usual, isomorphism between 
Hilbert spaces will be treated as equality):
\begin{equation}\label{U1-dec}
  \Uone=\int_{I}d\ra\Ua \hsc \Ua=(\chi_{\pma}\La)(\Pc(\pma))\uparrow\Pc
\end{equation}
and the representation space is:
\begin{equation}\label{H1-dec}
  \Hone=\int_{I}^{\oplus}d\ra\Ha \hsc \Ha=\Lsq_{\mma}(\Tma,\Hc(\La))
\end{equation}
where $I$ is an index set, $\rho$ is a measure on $I$ (determined
by $\Uone$ up to equivalence) and for each $\al\in I$,
\begin{description}
  \item[i.] $\Tma$ (the ``$\ma$-mass shell'') is an orbit of the ``Lorentz
    group'' $\Lc\so={\cal O}\so(r,s)$, in the ``momentum space'' $\T$,
    and $\mma$ is the non-trivial $\Lc\so$-invariant Radon measure on
    $\Tma$ (unique, up to a multiplicative constant); the non-triviality
    and the $\Lc$-invariance of $\mma$ imply that it is
    non-degenerate (it does not vanish on open sets);
  \item[ii.] $\pma$ is the representative of $\Tma$
    (as chosen in appendix B, \eg, for the time-like orbits of
    ${\cal O}(r,1),\;p=(p,0,\ldots,0)$, which in 4 dimensions
    corresponds to the rest frame);
  \item[iii.] $\La$ is an irreducible continuous unitary representation
    of the little group $\Lc(\pma)$ of $\pma$ (where $\Lc$ is the
    universal covering group of $\Lc\so$) in the (complex separable)
    Hilbert space $\Hc(\La)$;
  \item[iv.] $\Ha$ is the space of $\mma$-square-integrable functions on
    $\Tma$, taking values in $\Hc(\La)$;
  \item[v.] $\Ua$ is the irreducible (continuous and unitary)
    representation of $\Pc$ in $\Ha$, induced by
    $\La$:
    \begin{equation}\label{Uaf}
      [\Ua(\Lm,a)f](p)=e^{ip\cdot a}\La(\Dl(\Lm,p))f(\Lm^{-1}p)
    \end{equation}
    where
    \[ \Dl(\Lm,p)=\Lmp^{-1}\Lm\Lm_{\Lm^{-1}p} \]
    and $\forall p\in\Tma$, $\Lmp$ is in $\Lc$ and satisfies
    $\Lmp\pma=p$; thus $\Dl(\Lm,p)\in\Lc(\pma)$.
\end{description}

We want to treat the elements of $\Hone$ as vector-valued functions on
$\T$. For this, we assume that the order of the $d\ra$ integration can
be arranged to be
\begin{equation}
  \int_{I}d\ra\ldots=\int_{\M}d\mh(m)\int_{I(m)}d\ra\ldots
\end{equation}
where $\mh$ is a measure on a set of orbits $\{\Tm|m\in\M\}$ and for
each $m\in\M$,\linebreak $I(m):=\{\al\in I|\ma=m\}$ is the set of
indices of all
irreducible components of $\Uone$ with the ``mass'' $m$. (This
assumption should be satisfied if $I$ is not pathological.) In this
case we have (compare to (\ref{H1-dec})):
\begin{equation}\label{H1Hm}
  \Hone=\int_{\M}^{\oplus}d\mh(m)\int_{\Tm}^{\oplus}d\mm(p)\Hc(m) \hsc
  \Hc(m):=\int_{I(m)}^{\oplus}d\ra\Hc(\La)
\end{equation}
and $\Uone$ gets the form (compare to (\ref{Uaf})): for $p\in\Tm$,
\begin{equation}\label{Uone-def}
  [\Uone(\Lm,a)f](p)=e^{ip\cdot a}\Lp(\Dl(\Lm,p))f(\Lm^{-1}p)
\end{equation}
where $\Lp=\int_{I(m)}d\ra\La$ is the (reducible) unitary
representation of $\Lc(\psm)$ in $\Hc(m)$.

The momentum support of the elements of $\Hone$ is restricted to
$\TF:=\bigcup_{m\in\M}\Tm$. This will be called ``the (one particle)
physical region in $\T$'' and it is the spectrum of the momentum
operator in $\Hone$.

\subsection{Representing $\Hs$ as a Function Space}
\label{Func-Space}

The next step is to extend $\Uone$ from $\Pc$ to $\AP$, the Lie algebra
of $\Pc$. This is done by identifying the elements of $\Pc$ and $\UP$,
the universal enveloping algebra of $\AP$, as distributions on $\Pc$
with compact support, and then defining a representation of $\E'(\Pc)$,
the space of all such distributions. In this procedure, we follow
Antoine \cite{Antoine2} and it is described in appendix B.2
(refer also to appendix A for notation and terminology). The results
can be summarized as follows:
\begin{quote}
  If
  \begin{description}
    \item[C1.] $(\Pa,\Hc(\La),\Pa')$ is a Gel'fand triple:
      \begin{description}
        \item[(a)] $\Pa$ is a complete nuclear space, embedded in
          $\Hc(\La)$ densely and continuously,
        \item[(b)] $\Pa'$ is the strong dual of $\Pa$,
      \end{description}
    \item[C2.] the restriction of $\La$ to $\Pa$ is a smooth
      representation of $\Lc(\pma)$ by continuous operators in $\Pa$:
      \begin{description}
        \item[(a)] $\Pa$ is $\La$-invariant,
        \item[(b)] for each $\Dl\in\Lc(\pma),\;\La(\Dl)$ is a continuous
          operator in $\Pa$,
        \item[(c)] for each $\psi\in\Pa$, the function
          $\Dl\mapsto\La(\Dl)\psi$ is a smooth function from
          $\Lc(\pma)$ to $\Hc(\La)$,
      \end{description}
    \item[C3.] the map $p\mapsto\Lmp$ is smooth on $\Tma$,
  \end{description}
  then
  \begin{description}
    \item[R1.] the triple $(\Fa,\Ha,\Fa')$, where $\Fa$ is defined by
      \begin{equation} \Fa:=\D(\Tma;\Pa), \end{equation}
      is a Gel'fand triple, having properties (1a) and (1b) of
      $(\Pa,\Hc(\La),\Pa')$,
    \item[R2.] the restriction of $\Ua$ to $\Fa$ is a smooth representation
      of $\Pc$ by continuous operators in $\Fa$ (\ie $\Ua$ has the
      properties (2a), (2b), and (2c) of $\La$),
    \item[R3.] in $\Fa$, $\Ua$ is naturally extended to a continuous
      *-representation of $\UP$ (actually, of all of $\E'(\Pc)$) by
      continuous operators (the elements of $\UP$ being represented by
      differential operators on $\Tma$):
        \[ \Ua(A)\Ua(B)=\Ua(AB) \hsc \forall A,B\in\UP \]
        \[ \Ua(A\dg)\subset[\Ua(A)]^{*} \hsc \forall A\in\UP \]
      so if $A$ is symmetric $(A\dg=A)$ then $\Ua$ is Hermitian.
    \item[R4.] if $A\in\UP$ satisfies the Nelson-Stinespring criterion ($A$
      is elliptic -- when considered as a differential operator on
      $\E(\Pc)$ -- or there exists $B\in\UP$ symmetric and elliptic for
      which $[\Ua(A),\Ua(B\dg B)]=0$ on $\Fa$) then
        \[ \bbar{\Ua(A\dg)}=[\Ua(A)]^{*} \]
      ( $\bbar{\Ua(A\dg)}$ is the closure of $\Ua(A\dg)$ in $\Ha$) so
      if $A$ is symmetric then $\Ua(A)$ is essentially self adjoint 
      \cite{Nel-Sti},
    \item[R5.] if $A,B\in\UP$ are represented by essentially self adjoint
      operators $\Ua(A)$, $\Ua(B)$ then:
      \begin{equation}
        \bbar{\Ua(A)}\mbox{ and }\bbar{\Ua(B)}\mbox{ strongly
          commute }\Longleftrightarrow\;[A,B]=0.
      \end{equation}
  \end{description}
\end{quote}

In this situation (\ie under the conditions C1 and C2) it is
possible to represent $\Hone$ as a space of complex-valued
functions. Let $P={\Pm}$ be the ``momentum'' operator (the generator of
translations: $i\Pm=\partial/\partial a_{\mu}|_{a=0}$).
>From eq. (\ref{Uaf}) one gets that
  \[ \forall\f\in\Fa \hsc [\Ua(\Pm)\f](p)=p^{\mu}\f(p) \]
so $P$ is diagonal in $\Fa$ (it also demonstrates all the results R3,
R4 and R5 ). Using results R4 and R5 and the irreducibility of
$\Ua$, we choose a (finite) abelian set $J=\{J_{i}\}$ of symmetric
elements of $\UP$ such that for each $\al\in I$,
$\{\bbar{\Ua(P)},\bbar{\Ua(J)}\}$ is a complete system of strongly
commuting self adjoint operators and we diagonalize it, using
von~Neumann's complete spectral theorem \cite[p. 54]{Maurin}. This
diagonalization will only affect $\Hc(\La)$ (which is, in a generalized
sense, an eigenspace of $P$) so we get
\beq\label{HcLa}  \Hc(\La)\simeq\Lsq_{\na}(\sa)  \eeq
where $\sa$ is the spectrum of $\bbar{\Ua(J)}$ and $\na$ is a spectral
measure on $\sa$. Therefore:
\begin{equation}
  \Ha=\Lsq_{\mma}(\Tma;\Hc(\La))
     \simeq\Lsq_{\mma\times\na}(\Tma\times\sa)
\end{equation}
and for each $\f\in\Fa(\subset\Ha),\;p\in\Tma,\;\lm\in\sa$:
\begin{eqnarray}
  {}[\Ua(\Pm)\f](p,\lm)&=&p^{\mu}\f(p,\lm)\\
  {}[\Ua(J_{i})\f](p,\lm)&=&\lm_{i}\f(p,\lm).
\end{eqnarray}
Turning now to $\Hone$, we have
\begin{equation}
  \Hc(m)=\int_{I(m)}^{\oplus}d\ra\Hc(\La)=\Lsq_{\rho_{m}}(\Om(m))
\end{equation}
where
\[ \Om(m):=\{[\al\lm]|\al\in I(m),\;\lm\in\sa\} \]
and
\[ \int_{\Om(m)}d\rho_{m}(\al,\lm)\ldots:=
     \int_{I(m)}d\ra\int_{\sa}d\na(\lm)\ldots               \]
thus (compare to (\ref{H1Hm})):
\begin{equation}\label{H1Om} \Hone=\Lsq_{\mu}(\Om) \end{equation}
where
\begin{eqnarray*}
  \Om&:=&\{(p,\lm,\al)|m\in\M,\;p\in\Tm,\;[\al\lm]\in\Om(m)\}\\
    &=&\{(p,\lm,\al)|\al\in I,\;p\in\Tma,\;\lm\in\sa\}
\end{eqnarray*}
and
\[ \int_{\Om}d\mu(p,\lm,\al)\ldots\:=
     \int_{\M}d\mh(m)\int_{\Tm}d\mm(p)\int_{I(m)}d\ra
     \int_{\sa}d\na(\lm)\ldots\;.                            \]

The $n$-particle space is, according to (\ref{Hns}) \cite{eqHn}
\beq\label{Hn}
  \Hn=\bbar{\bigotimes_{1}\n\Lsq_{\mu}(\Om)}
  =\Lsq_{\mu\n}(\Om\n)
\eeq
where
\beq\label{Omn}
  \Om\n:=\Om\times\cdots\times\Om\mbox{ ($n$ factors)}
\eeq
and the measure $\mu\n$ is defined by
\beq\label{mun}
  \int_{\Om\n}d\mu\n((p,\lm,\al)\n)\ldots:=
    \int_{\Om}d\mu(p_1,\lm_1,\al_1)\ldots
    \int_{\Om}d\mu(p_n,\lm_n,\al_n)\ldots\;.
\eeq

\newsection{The Complete Set of Plane Waves}

At this point we introduce further assumptions about the signature
of space-time and the spectrum of particle types:
\begin{quote}
  \bf Assumptions:\em
  \begin{description}
    \item[3.1] $\Lc\so={\cal O}\so(r,1)$, $r\geq3$ and a momentum $p$ in the 
      physical region $\TF$ is time-like ($p_{\mu}p^{\mu}>0$) and in the 
      forward light cone ($E>0$) (thus $m$ can be recognized as a generalized 
      mass $m=\sqrt{p_{\mu}p^{\mu}}$; see Appendix B for notation).
    \item[3.2] $I$ is countable (thus $\rho$ is a purely atomic measure) and
      $\{\ma\}_{\al\in I}$ is a non-decreasing sequence.
  \end{description}
\end{quote}
>From now on, we also assume that the map $p\mapsto\Lmp$ was chosen to
be smooth and $\Lm_{\psm}=1,\;\forall m\in\M$.

\subsection{The One-Particle Space}
\label{PW1}

Assumption 3.1 means that $\Lc\so(\psm)={\cal O}_{0}(r)$ which is a
compact group. Since \mbox{$r\neq2$}, it is at most doubly covered by
$\Lc(\psm)=\bbar{\cal O}\so(r)$, therefore $\Lc(\psm)$ is also compact,
so all its irreducible representations are finite dimensional:
$\forall\al\in I$, \mbox{$\dim\Hc(\La)<\infty$}. This implies that:
\begin{description}
  \item[(1a)] $\Hc(\La)$ is a nuclear space \cite[p. 520]{Treves};
  \item[(1b)] since $\La$ is continuous, it is smooth (even analytic
    \cite[p. 322]{Barut});
\end{description}
thus all the requirements for the extension of $\Ua$ to $\UP$ are
satisfied with $\Pa=\Hc(\La)$.

Assumption 3.1 also implies that all the ray representations of $\Pc$
are (equivalent to) true representations \cite{Barg-ray}, so
assumption 2.2 (in Section \ref{Psymm}) is satisfied.

Assumption 3.2 means that:
\begin{description}
  \item[(2a)] $\M$ (the mass spectrum) is a countable set of discrete
    points in $(0,\infty)$ (with no accumulation points). Thus
    \begin{equation}
      \Hone=\bigoplus_{m\in\M}\Lsq_{\mm}(\Tm,\Hc(m));
    \end{equation}
  \item[(2b)] For each $m\in\M$, $I(m)$ (the set of indices of all
    irreducible components of $\Uone$ with the ``mass'' $m$) is finite:
    \begin{equation}
      \Hc(m)=\bigoplus_{\al\in I(m)}\Hc(\La),\;
      \Lp=\bigoplus_{\al\in I(m)}\La,
      \mbox{ where }m=\sqrt{p_{\mu}p^{\mu}}
    \end{equation}
  \item[(2c)] $\Uone$ is a direct sum of irreducible representations:
    \begin{equation}
      \Hone=\bigoplus_{\al\in I}\Ha,\;\Uone=\bigoplus_{\al\in I}\Ua
    \end{equation}
\end{description}

Combining both assumptions, we get:
\begin{description}
  \item[(3a)] $\Hc(m)$ is finite-dimensional and since it is (identified
    with) a space of functions, we have:
    \begin{equation}
      \Hc(m)={\bf C}^{N(m)} \hs\hs (\;N(m):=\dim\Hc(m)\;)
    \end{equation}
    thus operators in $\Hc(m)$ (\eg $\Lp$) are matrices.
  \item[(3b)] $\Om$ (defined in eq. (\ref{H1Om})) is a countable union
    of orbits. Since each orbit is a separable smooth manifold, so is
    $\Om$.

    The measure $\mu$ on $\Om$ takes the form:
    \begin{eqnarray}
      \int_{\Om}d\mu(p,\lm,\al)\ldots & = &
        \sum_{\al\in I}\int_{\Tma}d\mma(p)\sum_{\lm\in\sa}\ldots \\
      & = & \sum_{m\in\M}\int_{\Tm}d\mm(p)\sum_{[\al\lm]\in\Om(m)}
                                                      \ldots \nonumber
    \end{eqnarray}
    (compare to eq. (\ref{H1Om})) and since on each orbit $\Tm$, $\mm$
    is a non degenerate \mbox{($\Lc$-invariant)} Radon measure, so is
    $\mu$ on $\Om$.
\end{description}
Writing everything as a function on the momentum space, it is frequently
convenient to write $\Hc(p),N(p),\Om(p)$ and $I(p)$ instead of
$\Hc(m)$ etc., where \mbox{$m=\sqrt{p_{\mu}p^{\mu}}$}.

Now we are able to identify the structure of the space of states
required by the Dirac formalism (see appendix A.2). Recalling
that $\Hone=\Lsq_{\mu}(\Om)$, We define:
\begin{equation} \Fone:=\D(\Om). \end{equation}
Since $I$ is countable, it can be chosen to be a set of natural
numbers, so the operator $\hat{\al}$ defined by
\begin{equation}
  [\hat{\al}f](p,\lm,\al):=\al f(p,\lm,\al),\;\forall f\in\Hone
\end{equation}
is self adjoint and diagonal. Recognizing that $\Lsq_{\mu}(\Om)$ is
the spectral decomposition of $\Hone$ with respect to the system
$\{\hat{\al},\bbar{\Uone(\Pm)},\bbar{\Uone(J_{i})}\}$, one obtains
that this is a complete system of strongly commuting self adjoint
operators. Next, we recognize that
\begin{equation} \Fone=\sum_{\al\in I}\Fa \end{equation}
(a topological direct sum of locally convex spaces: the set of {\em
finite} sums of elements of $\{\Fa\}_{\al\in I}$ \cite[p. 515]{Treves})
so (according to result R3 of Section \ref{Func-Space}) all the above
operators, when restricted to $\Fone$ are continuous operators in $\Fone$.
Thus $(\Fone,\Hone,\Phi^{(1)'})$ is a Gel'fand triple with all the
properties required, so all the results described in appendix A
apply here. In particular, the set
$\{<p,\lm,\al|\}_{(p,\lm,\al)\in\Om}$, defined ($\mu$-almost
everywhere) by
\begin{equation}
  <p,\lm,\al|\f):=\f(p,\lm,\al),\;\forall\f\in\Fone
\end{equation}
is a ``complete orthonormal'' system of eigenbras of
$\{\hat{\al}',\bbar{\Uone(\Pm)'},\bbar{\Uone(J_{i})'}\}$ (in the sense
explained in appendix A.4).

In the following, to simplify notation, the indices $[\al\lm]$ will be
usually omitted, making the summations implicit (``matrix
multiplication''); $<p,\lm,\al|$ and $d\mu(p,\lm,\al)$ are abbreviated
by $<p|$ and $d\mu(p)$.

\subsection{The Total Space}

The $n$-particle space is (see (\ref{Hn}))
\[ \Hn=\Lsq_{\mu\n}(\Om\n) \]
where $\Om\n$ is a separable smooth manifold and $\mu\n$ is a non
degenerate Radon measure (see result 3b in the previous subsection).
This suggests that the natural choice for $\Fn$ is
\beq  \Fn:=\D(\Om\n),  \eeq
obtaining a Gel'fand triple $(\Fn,\Hn,\Fnd)$.

For each $p\n=(p_{1},\ldots,p_{n})\in\Om\n$ we define:
\beq  <p\n|:=<p_{1}|\otimes\cdots\otimes<p_{n}|  \eeq
and since $\bigotimes_{1}\n\D'(\Om)$ is a (dense) subspace of
$\D'(\Om\n)$ \cite[p. 417]{Treves}, we get
\beq  <p\n|\in\Fnd  \eeq
with the action (for $\mu\n$-almost all $p\n\in\Om\n$)
\beq  <p\n|\f)=\f(p\n),\;\forall\f\in\Fn.  \eeq
For $\f,\psi\in\bigotimes_{1}\n\D(\Om)$, we have
\begin{eqnarray}
  (\psi|\f) & = & (\psi,\f)_{\Hn}
      = \prod_{i=1}\n(\psi_{i},\f_{i})_{\Hone}   \label{Par-n}\\
    & = & \prod_{i=1}\n\int_{\Om}d\mu(p_{i})
            \bbar{\psi_{i}(p_{i})}\f_{i}(p_{i})      \nonumber\\
    & = & \int_{\Om\n}d\mu\n(p\n)(\psi|p\n><p\n|\f). \nonumber\\
\end{eqnarray}
This is exactly the Parseval equality (compare to eq. (\ref{pars}))
which means that $\{<p\n|\}$ is a complete orthonormal system of
bras in the sense that the operator
\beq  I\n:=\int_{\Om\n}d\mu\n(p\n)|p\n><p\n|  \eeq
is the embedding of $\Fn$ into $\Fnd$ and plays the role of the
identity operator in the Dirac formalism \cite{Ires}.

Next, we define
\beq
  \Hc:=\bigoplus_{n=1}^{\infty}\Hn,\;
  \Phi:=\sum_{n=1}^{\infty}\Fn
\eeq
obtaining a Gel'fand triple $(\Phi,\Hc,\Phi')$. Since $\Fn$ is a
closed subspace of $\Phi$, $\Fnd$ is a quotient space of $\Phi'$ and can be
naturally identified  as a (closed) subspace of $\Phi'$, by defining
\beq  <\Phi^{(m)'}|\Fn>\propto\dl_{mn}.  \eeq
Therefore we get
\beq  \Phi'=\prod_{n=1}^{\infty}\Fnd,  \eeq
(the set of {\em arbitrary infinite} sums) 
and the embedding of $\Phi$ into $\Phi'$ is
\beq\label{Iexp}
  I:=\sum_{n=1}^{\infty}I\n
    =\sum_{n=1}^{\infty}\int_{\Om\n}d\mu\n(p\n)|p\n><p\n|.
\eeq
Finally, we define
\beq  \Fs:=\Phi\cap\Hs,\;\Fns:=\Fn\cap\Hns  \eeq
obtaining the Gel'fand triples $(\Fs,\Hs,\Fs')$ and
$(\forall n)\;(\Fns,\Hns,\Fnsd)$, equipped with the complete systems of
bras defined above.

\subsection{Matrix Elements}
\label{Matrix}

Let $A\in L^{\times}(\Phi;\Phi')$. As stated in appendix A.3, this space 
contains all the spaces 
$L(\Phi),L(\Hc),L(\Phi;\Hc),L^{\times}(\Hc;\Phi')$, and $L(\F')$
where ``$L$'' denotes ``linear'' and ``$L^{\times}$'' -- ``antilinear''.
Moreover, $\Fs$ is a closed subspace of $\Phi$ so if
$A\in L^{\times}(\Fs;\Fs')$, it can always be extended continuously
to all of $\Phi$ by defining $Af:=0$ for each $f\in\Hc$ which is
orthogonal to $\Fs$. This means that $L^{\times}(\Fs;\Fs')$, and all
the corresponding spaces of continuous mappings, are also embedded
naturally in $L^{\times}(\Phi;\Phi')$.

To define the matrix elements of $A$, we decompose it to ``elements''
acting between states of definite number of particles.
Let $\tau\n$ be the natural (continuous isomorphic) embedding of $\Fn$
in $\Phi$, and $\tau^{n'}$, the dual mapping (this is the natural
projection of $\Phi'$ on $\Fnd$). We define
\beq  A\mn:=\tau^{m'}A\tau\n\;(\in L^{\times}(\Fn;\Phi^{(m)'})\;)  \eeq
obtaining, for each $\f,\psi\in\Phi$
\beq\label{el1} (\psi|A\f>=\sum_{m,n}(\psi\m|A\mn\f\n>  \eeq
(with the notation $\f=\sum_{n}\f\n,\;\f\n\in\Fn$ and the same for
$\psi$).

Now we can apply the kernel theorem (see appendix A.4), obtaining
\beq\label{el2}
  (\psi\m|A\mn\f\n>=\int_{\Om\m\times\Om\n}d\mu\m(q\m)d\mu\n(p\n)
    \bbar{\psi\m(q\m)}A\mn(q\m,p\n)\f\n(p\n)
\eeq
where $<A\mn>\in\D'(\Om\m\times\Om\n)$ is the kernel that corresponds
to $A\mn$. Denoting
\beq  <q\m|A|p\n>:=A\mn(q\m,p\n)  \eeq
we obtain (combining (\ref{el1}) and (\ref{el2}) )
\beq
  (\psi|A\f>=\sum_{m,n}\int_{\Om\m\times\Om\n}d\mu\m(q\m)d\mu\n(p\n)
    (\psi|q\m><q\m|A|p\n><p\n|\f)
\eeq
so we see that , as in the general formalism, the expression
(\ref{Iexp}) for $I$ can be formally inserted between the factors of
$(\psi|A|\f>$.

\newsection{Applications}

In this Section we illustrate the formalism derived above by rederiving
familiar relations and formulas using the new language. The resemblance to
the original Dirac formalism is apparent but unlike the original formalism,
in the present formulation all the expressions have a well defined
meaning.

\subsection{The Representation $\Uone$}

>From eq. (\ref{Uone-def}) we obtain
\begin{eqnarray}
  <\!p|\Uone(\Lm,a)|\f) & = & [\Uone(\Lm,a)\f](p)=             \\
    & = & e^{ip\cdot a}\Lp(\Dl(\Lm,p))<\!\Lm\inv p|\f)  \nonumber\\
\end{eqnarray}
so the action of $\Uone$ on the base vectors is
\beq <\!p|\Uone(\Lm,a)=e^{ip\cdot a}\Lp(\Dl(\Lm,p))<\!\Lm\inv p|. \eeq
(here $<p|$ is considered as a column vector --- because when acting on
$|\f)$, it produces a column vector in $\Hc(p)$ --- thus $|p>$ is
recognized as a row vector).

It is more customary to write the expression for $\Uone(\Lm,a)|p>$
and this is equal to $(<p|\Uone(\Lm,a)^{*})\dg$, where ``$\dg$''
denotes the matrix conjugation in $\Hc(p)$ (and not the adjoint
defined in appendix A.4). $\Uone(\Lm,a)$ is unitary and
$(\Lm,a)\inv=(\Lm\inv,-\Lm\inv a)$ (see appendix B), so
\begin{eqnarray*}
  \Uone(\Lm,a)|p> & = & {}[<p|\Uone(\Lm\inv,-\Lm\inv a)]\dg=       \\
    & = & {}[e^{-ip\cdot \Lm\inv a}\Lp(\Dl(\Lm\inv,p))<\!\Lm p|]\dg= \\
    & = & |\Lm p\!>e^{ip\cdot \Lm\inv a}\Lp(\Dl(\Lm\inv,p))\dg
\end{eqnarray*}
$\Lp$ is a unitary representation and
\[  \Dl(\Lm\inv,p)\inv=(\Lmp\inv\Lm\inv\Lm_{\Lm p})\inv
      =\Lm_{\Lm p}\inv\Lm\Lmp=\Dl(\Lm,\Lm p)             \]
so (since also $p\cdot\Lm\inv a=a\cdot\Lm p$)
\beq \Uone(\Lm,a)|\!p>=|\Lm p\!>e^{ia\cdot\Lm p}\Lp(\Dl(\Lm,\Lm p)) \eeq
and explicitly, with components:
\beq
  \Uone(\Lm,a)|p,\lm,\al\!>=
    e^{ia\cdot\Lm p}\La(\Dl(\Lm,\Lm p))_{\lm'\lm}|\Lm p,\lm',\al>.
\eeq
This also implies that the matrix elements of $\Uone$ are
\beq
  <\!p'|\Uone(\Lm,a)|p\!>=
    e^{ia\cdot\Lm p}\Lp(\Dl(\Lm,\Lm p))\dl_{\mu}(p'-\Lm p)
\eeq
or, in components:
\beq
  <\!p',\lm',\al'|\Uone(\Lm,a)|p,\lm,\al\!>=e^{ia\cdot\Lm p}
    \dl_{\al'\al}\La(\Dl(\Lm,\Lm p))_{\lm'\lm}\dl_{\mu}(p'-\Lm p)
\eeq
and indeed ,one gets
\beq
  <p|\Uone(\Lm,a)|\f)=
    \int_{\Om}d\mu(p)<p|\Uone(\Lm,a)|p'><p'|\f).
\eeq

\subsection{Generators of Symmetry}

Let $g(t)$ be a one-parameter symmetry group of $S$. As such, it is
represented in $\Hone$ by a unitary representation $\Uone$. {\em The
generator $A_g$ of $\Uone(g(t))$} is defined by
\beq
  (\psi|\Ag\f\!>:=\frac{1}{i}\der{t}(\psi,\Uone(g(t))\f)|_{t=0},\;
  \forall\f,\psi\in\Fone
\eeq
and it is assumed to be an element of $L^{\times}(\Fone;\Foned)$.
(this means, in particular, that the derivative exists for each
$\f,\psi$ and the function thus obtained is continuous in $\psi$
and $\f$) \cite{Stone}.
If $g(t)$ is a subgroup of $\Pc$, this assumption is certainly
satisfied: $\Fone$ was intentionally constructed to make a generator of
such a group a continuous operator in $\Fone$.

The unitarity of $\Uone$:
\beq  \Uone(g(t))^{*}=\Uone(g(t)\inv)=\Uone(g(-t))  \eeq
implies that $\Ag$ is self adjoint (as an
element of $L^{\times}(\Fone;\Foned)$: $A\dg=A$):
\begin{eqnarray}
  (\psi|\Ag\dg\f\!> & = & \bbar{(\f|\Ag\psi\!>}=
    \bbar{\frac{1}{i}\der{t}(\f,\Uone(g(t))\psi)|_{t=0}}=      \\
  & = & -\frac{1}{i}\der{t}(\psi,\Uone(g(t))^{*}\f)|_{t=0}=  \nonumber\\
  & = & -\frac{1}{i}\der{t}(\psi,\Uone(g(-t))\f)|_{t=0}=  \nonumber\\
  & = & \frac{1}{i}\der{t}(\psi,\Uone(g(t))\f)|_{t=0}=
    (\psi|\Ag\f\!>                                   \nonumber
\end{eqnarray}

Next we turn to multi-particle states. $\Ag\mn$ is defined naturally
by:
\beq
  (\psi\m|\Ag\mn\f\n\!>:=\frac{1}{i}\der{t}
  (\psi\m,U(g(t))\f\n)|_{t=0},\;\forall\psi\m\in\F\ms,\f\in\Fn.
\eeq
Since $U$ doesn't change the number of particles, neither does $\Ag$,
so
\beq  \Ag\mn=0,\;\forall m\neq n.  \eeq
For $\psi=\bigotimes_{1}\n\psi_{i},\;\f=\bigotimes_{1}\n\f_{i}$ in
$\bigotimes_{1}\n\Fone$ we have
\begin{eqnarray}
  (\psi|\Ag\ns\f\!> & = & \frac{1}{i}\der{t}
    [\prod_{i=1}\n(\psi_{i},\Uone(g(t))\f_{i})]|_{t=0}  \\
  & = & \sum_{i=1}\n(\psi_{i}|\Ag\f_{i}\!>
    \prod_{j\neq i}(\psi_{j},\f_{j})  \nonumber
\end{eqnarray}
so at least between elements of $\bigotimes_{1}\n\Fone$:
\beq\label{Agn}
  \Ag\ns=(\Ag\otimes I\otimes\cdots\otimes I)
        +(I\otimes\Ag\otimes\cdots\otimes I)
        +\cdots+(I\otimes I\otimes\cdots\otimes \Ag)
\eeq
(to be extended to arbitrary elements of $\Fn$, $\Ag\ns$ must be
continuous).

Finally, $g(t)$, being a group of symmetries of the S-matrix $S$,
satisfies $[U(g(t)),S]=0$. This implies that for each
$\f,\psi\in\F$,
\beq  (S^{*}\psi,U(g(t))\f)=\bbar{(S\f,U(g(-t))\psi)}.  \eeq
Differentiating, one gets (when $S^{*}\psi,S\f\in\F$)
\beq  (S^{*}\psi|\Ag\f\!>=<\!\Ag\psi|S\f).  \eeq
In particular, if $S$ and $\Ag$ are operators in $\F$ then
\beq\label{com}  [\Ag,S]=0\mbox{ in }\F.  \eeq
Also, if $\Ag$ is a {\em continuous operator in} $\F$ (\eg a generator
of $\Pc$) then (\ref{com}) holds, with the commutators defined to be
\beq  [\Ag,S]=\Ag'S-S\Ag  \eeq
where $\Ag'$ is the dual of $\Ag$ and $S$ is considered as an operator
from $\F$ to $\F'$.

\subsection{Scattering Amplitudes}

We have assumed (see the beginning of Section 2) that the S-matrix $S$ is 
unitary. This implies that it is continuous in $\Hc_s$ and, therefore, can
be identified as an element of $L^{\times}(\F;\F')$. As such, it has a
corresponding kernel $<\!S\!>$ (more precisely, each element
$S\mn\in L(\Hn;\Hc\ms)$, as defined in Section \ref{Matrix}, has a
corresponding kernel $<\!S\mn\!>\in\D(\Om\m\times\Om\n)$ ). The momentum
operator $P$ is a generator of $\Pc$. Applying eq.\ (\ref{Agn}) we get
\beq
  [P\f](p_{1},\ldots,p_{n})=(p_{1}+\cdots+p_{n})\f(p_{1},\ldots,p_{n})
  ,\;\forall\f\in\bigotimes_{1}\n\Fone
\eeq
and applying eq.\ (\ref{com}), we get, for all
$\f\in\bigotimes_{1}\n\Fone$,
\begin{eqnarray}
  0 & = & ([P,S]\f)(q\m)=([P,S\mn]\f)(q\m)=       \\
    & = & \int d\mu\n(p\n)(\sum_{1}\m q_{j}-\sum_{1}\n p_{i})
          <\!q\m|S|p\n\!>\f(p\n)d\mu\n(p\n),  \nonumber\\
\end{eqnarray}
and hence
\beq\label{cons}
  (\sum_{1}\m q_{j}-\sum_{1}\n p_{i})<\!q\m|S|p\n\!>=0;
\eeq
obviously this holds also for $S-I$ replacing $S$, where $I$ is
the identity operator in $\Hc$. This implies that $<\!S-I\!>$
is of the form \cite[vol. 1]{Gel'fand}
\beq\label{SIel}
  <\!S-I\!>=-i(2\pi)^{d}\dl^{d}(\sum_{1}\m q_{j}-\sum_{1}\n p_{i})<\!T\!>
\eeq
where $d=r+1$ is the dimension of the momentum space, $-i(2\pi)^{d}$
is a conventional normalization factor and $<\!T\!>$ is a generalized
function on the submanifold of $\Om\m\times\Om\n$ defined by the
constraint
\beq  \sum_{1}\m q_{j}-\sum_{1}\n p_{i}=0.  \eeq
(this is the precise formulation of energy-momentum conservation) so
finally:
\beq
  <\!S\mn\!>=<\!I\mn\!>
    -i(2\pi)^{d}\dl^{d}(\sum_{1}\m q_{j}-\sum_{1}\n p_{i})<\!T\mn\!>.
\eeq
The (generalized) values of $<\!T\!>$ are called ``scattering
amplitudes''.

\subsection{The Optical Theorem}

Since the indices $[\al\lm]$ are omitted, $<\!q\m|T|p\n\!>$ is a
(matrix) operator from \linebreak
$\Hc(p\n):=\bigotimes_{1}\n\Hc(p_{i})$ to $\Hc(q\m)$
(defined similarly) and the integration $d\mu\n(p\n)$ is actually over
$\TF\n\equiv\prod_{1}\n\TF$. In this Section, ``$\dg$'' denotes the
Hermitian conjugation of matrices. Bearing all this in mind, we now
show that the unitarity of $S$ leads to

\noindent{\bf The Optical Theorem:}
\begin{eqnarray}
  \lefteqn{<\!p\n|T|p\n\!>-<\!p\n|T|p\n\!>\dg=}  \\
  & & i\sum_{m=0}^{\infty}\int_{\TF\m}d\mu\m(q\m)(2\pi)^{d}\dl^{d}
      (\sum_{1}\m q_{j}-\sum_{1}\n p_{i})
      <\!p\n|T|q\m\!>\dg<\!q\m|T|p\n\!>   \nonumber
\end{eqnarray}
Proof:
\begin{quote}
  The unitarity of $S$ implies that:
  \beq  (I-S)^{*}(I-S)=(I-S)+(I-S)^{*}.  \eeq
  Writing the corresponding equation for kernels (using the
  expression (\ref{SIel}) for $<\!I-S\!>$) we get
  \begin{eqnarray}
    0 & = & <\!p\n|[(I-S)+(I-S)^{*}-(I-S)^{*}(I-S)]|r^{l}\!>=  \\
      & = & i(2\pi)^{d}\dl^{d}(\sum_{1}\n p_{i}-\sum_{1}^{l}r_{k})
            \times  \nonumber\\
      &   & \times\{<\!p\n|T|r^{l}\!>-<\!p\n|T|r^{l}\!>\dg-
            \sum_{m=1}^{\infty}\int_{\TF\m}d\mu\m(q\m)  \nonumber\\
      &   & i(2\pi)^{d}\dl^{d}(\sum_{1}\m q_{j}-\sum_{1}\n p_{i})
            <\!p\n|T|q\m\!>\dg<\!q\m|T|r^{l}\!>\}  \nonumber
  \end{eqnarray}
  so for $\sum_{1}\n p_{i}=\sum_{1}^{l}r_{k}$, the expression in \{\}
  must vanish. In particular, for $r^{l}=p\n$. \qed
\end{quote}

\newsection{Comments and Supplements}
In this Section we discuss the assumptions used for the construction,
emphasizing the prospects for relaxing some of them.

\subsection{Other Signatures and Orbits}
In Section 3 we assumed signatures of the type $(r,1)$
and representations with momentum support in the forward light cone. All
this was needed to assure the compactness of the little group, which
implies that its irreducible representation spaces $\{\Hc(\La)\}$ are
finite-dimensional. This plays a key role in the
construction of the Gel'fand triple. If $\Hc(\La)$ is of infinite
dimension, it is never nuclear \cite[p. 520]{Treves} and $\La$ is, in
general, not smooth. Therefore the choice $\Pa:=\{\Hc(\La)\}$ made in
Section \ref{PW1} cannot satisfy in this case the requirements of
Section \ref{Func-Space} for the construction of a Gel'fand triple
suitable for $\AP$.
If $\La$ itself is induced by a finite dimensional representation, the
space $\Pa$ can be constructed by the same procedure described in
Section 2.3 for $\Fa$. Using this procedure, $\Fa$ can be constructed
for any representation which can be built by a sequence of inductions,
starting with a finite-dimensional representation.

The infinite dimension of $\{\Hc(\La)\}$ may cause another complication.
In this case, the spectrum $\sa$ (of the operator $J$ used to
represent $\{\Hc(\La)\}$ as a space of functions -- eq. (\ref{HcLa}))
is not necessarily discrete. If it is continuous, $\Om$ (defined in
eq. (\ref{H1Om})) is not a countable union of orbits so to consider
$\Om$ as a smooth separable manifold, one must include a differential
structure on $\sa$; this must be taken into account when checking the
smoothness of functions on $\Om$. If the spectrum is mixed, $\Om$ is
a union of manifolds of different dimension.

Finally, the choice of signature $(r,1)$ and momenta in the forward
light cone has also a physical significance. In this region $p^{0}$
is bounded from below (positive), thus suitable to be interpreted as
the energy. In any other case (except for the forward light-like
momenta in the case of signature $(r,1)$) the orbits are unbounded
in all directions, and therefore the canonical energy is not well
defined (recall that the energy is distinguished from other
components of the momentum by being positive and this in an invariant
--- and therefore well defined --- statement only in the case of signature
$(r,1)$).

\subsection{The Particle-Type Spectrum}

A particle type is identified with an irreducible representation of
$\Pc$. The assumption that it is a {\em true} representation (assumption
2.2 of Section \ref{Psymm}) is used in the
extension of the representation $U$ from $\Pc$ to its algebra $\AP$
(see appendix B.2). When $U$ is a genuine projective
representation (with a non-trivial phase), $\Ut$
(defined by eq. (\ref{UtTf})) is not a representation of $\E'(\Pc)$
(it doesn't conserve multiplication).

The discreteness of the set of particle types (assumption 3.2 of Section
3) was used (with the finiteness of dim$\Hc(\La)$) to identify $\Om$ as
a countable union of orbits, as discussed in the previous subsection.

\newsection{Conclusions}

In this work we constructed explicitly and rigorously a basis of
``plane-waves'' -- momentum (generalized) eigenstates -- for the space
of states used to describe relativistic scattering. We exploited the
assumed $\Pc\so$ symmetry, and used the theory of induced
representations and the structure of Gel'fand triples. To combine rigor
and clarity we used the a rigorized version of Dirac's ``bra'' and ``ket''
formalism. We develop this
formalism further and introduce a convenient notation which distinguishes
bra's $<\cdot|$, $|\cdot>$ from ket's $(\cdot|$, $|\cdot)$ \cite{Jauch}.
This notation made it possible to use the ``complete set of states'' to
decompose expressions into ``vector components'' and ``matrix
elements'' in almost the same flexibility as in the original
formalism. We demonstrate this flexibility in a few examples. A further
demonstration is given in \cite{Horwitz-Pelc}, where the construction
of the present work is used to prove an extension of the Coleman-Mandula
theorem.

\appendix
\renewcommand{\newsection}[1]{
 \vspace{10mm} \pagebreak[3]
 \addtocounter{section}{1}
 \setcounter{equation}{0}
 \setcounter{subsection}{0}
 \setcounter{paragraph}{0}
 \setcounter{equation}{0}
 \setcounter{figure}{0}
 \setcounter{table}{0}
 \addcontentsline{toc}{section}{
  Appendix \protect\numberline{\Alph{section}}{#1}}
 \begin{flushleft}
  {\large\bf Appendix \thesection. \hspace{5mm} #1}
 \end{flushleft}
 \nopagebreak}

\newsection{The Dirac Formalism}

\subsection{Conventional Terminology}

\subsubsection{Spaces of Operators}

(In this subsection, $E,F$ are topological vector spaces over the
complex field)
\begin{itemize}
  \item $L(E,F)$ : the space of continuous linear mappings from $E$ to
    $F$;
  \item $L^{\times}(E,F)$ : the space of continuous antilinear mappings
    from $E$ to $F$;

     ($L(E,F)$ and $L^{\times}(E,F)$ are naturally [antilinearily]
      isomorphic)
  \item $L(E):=L(E,E)$ and $L^{\times}(E):=L^{\times}(E,E)$;
  \item $E':=L(E;{\bf C})$ : ``the dual of $E$''; the space of
    continuous linear functionals on $E$; when endowed with the
    ``strong dual topology'', it is called ``the strong dual'';
  \item $\bar{E}':=L^{\times}(E,{\bf C})$.
\end{itemize}
When $E$ is reflexive (which means that it is the strong dual of its
strong dual) then $(\bar{E}')'$ is naturally (antilinearily) isomorphic
to $E$ and therefore is denoted by $\bar{E}$.

\subsubsection{Spaces of Functions}

(In this subsection, $\Om$ is a separable smooth (differentiable)
manifold, $\mu$ is a measure on $\Om$ and $\Hc,\Hc(x)$ are Hilbert
spaces)
\begin{itemize}
  \item $\int_{\Om}^{\oplus}d\mu(x)\Hc(x)$ : a direct integral of
    Hilbert spaces. An element of this space is a vector field
    \[  f:x\in\Om\mapsto f(x)\in\Hc(x).  \]
    This is a Hilbert space with respect to the inner product
    \[  (f,g):=\int_{\Om}d\mu(x)(f(x),g(x))_{\Hc(x)}  \]
    where $(\;,\;)_{\Hc(x)}$ is the inner product in $\Hc(x)$;
  \item $\Lsq_{\mu}(\Om;\Hc):=\int_{\Om}^{\oplus}d\mu(x)\Hc$ :
    the space of $\mu$-square-integrable functions from $\Om$ to $\Hc$;
  \item $\E(\Om;\Hc)$ : the space $C^{\infty}(\Om;\Hc)$ of smooth
    (infinitely differentiable) functions on $\Om$, with values in
    $\Hc$, equipped with the ``Schwartz topology'' (uniform
    convergence on every compact set in $\Om$ of the functions and all
    their derivatives);
  \item $\D(\Om;\Hc)$ : the space $C_{c}^{\infty}(\Om;\Hc)$ of those
    elements of $C^{\infty}(\Om;\Hc)$ that have compact support,
    equipped with the ``Schwartz topology''. (A sequence of functions
    $\varphi_{k}\in\Phi$ converges in this topology iff they have a
    common compact subset of $\Omega$ containing their supports and
    for each differential operator $D$, the sequence $\{D\varphi_{k}\}$
    converges uniformly \cite[p. 147]{Reed}). The elements of
    $\D(\Om;\Hc)$ are called ($\Hc$-valued) ``test functions on $\Om$''.

    These topologies were introduced by Schwartz \cite{Schwartz}
    and are nuclear \cite[p. 69]{Maurin} and complete \cite{Treves}.
    $\D(\Om;\Hc)$ is also reflexive \cite{Treves};
  \item $\D'(\Om;\Hc)$ : the strong dual of $\D(\Om;\Hc)$.
    Its elements are called ($\Hc$-valued) ``distributions on $\Om$'';
  \item $\E'(\Om;\Hc)$ : the strong dual of $\E(\Om;\Hc)$. Consists of
    those elements of $\D'(\Om;\Hc)$ that have compact support.
\end{itemize}
(When $\Hc=\bf C$, the field of complex numbers, this label is omitted.
\eg \linebreak $\Lsq_{\mu}(\Om,{\bf C})=\Lsq_{\mu}(\Om)$.)

\subsection{The Space of States}

The space of states is a Gel'fand triple (originally called ``rigged
Hilbert space'' by Gel'fand {\em et al.\ } \cite{Gel'fand}) --
a triplet $(\F,\Hc,\F')$ of topological vector spaces with the
following properties:
\begin{enumerate}
  \item $\Hc$ is a complex separable Hilbert space;
  \item $\F$ is a dense subspace of $\Hc$, equipped with a finer
    topology (more open sets; this is equivalent to the statement that
    the embedding of $\F$ in $\Hc$ is continuous);
  \item $\F'$ is the strong dual of $\F$ (\ie the topological dual,
    equipped with the strong dual topology \cite{Treves}).
\end{enumerate}
In the present context, it is further required that $\F$ be complete,
``nuclear'' \cite{Maurin,Treves} and reflexive ($\F$ is the strong
dual of $\F'$).

The space of states is equipped with a ``complete set of commuting
observables'' $\{A_i\}$: mutually strongly commuting self adjoint
operators which, when restricted to $\F$ are continuous in the topology 
of $\F$ \cite{CSCO}. According to the complete spectral theorem of von
Neumann \cite[p.54]{Maurin}, $\Hc$ is isomorphic to $\Lsq_{\mu}(\Om)$,
where $\Om$ is the combined spectrum of $\{A_i\}$. The observables are
so chosen that $\Om$ is a separable differentiable manifold (or a
discrete union of such manifolds), $\mu$ is
a non degenerate Radon measure on $\Om$ and $\F=\D(\Om)$.

\subsection{The Elements in the Formalism}

The types of vectors in $(\F,\Hc,\F')$:
\begin{itemize}
  \item {\bf ket vectors:} elements $|\f),|\psi),\ldots$ of $\F$ and
     elements $(\f|,(\psi|,\ldots$ of $\bbar{\F}$;
  \item {\bf normalizable vectors:} elements $f,g,\ldots$ of $\Hc$;
  \item {\bf bra vectors:} elements $<\xi|,<\zeta|,\ldots$ of $\F'$
    and elements $|\xi>,|\zeta>,\ldots$ of $\bbar{\F}'$.
\end{itemize}
There are three products
\begin{itemize}
  \item $(f,g)_{\Hc}$ is the inner product (linear in $g$) in $\Hc$
    (the subscript $\Hc$ is usually omitted);
  \item $<\xi|\f)$ is the dual action between $<\xi|$ and $|\f)$
    (the Dirac ``bracket'');
  \item $(\f|\xi>$ is the dual action between $(\f|$ and $|\xi>$
    (this would deserve the name ``ketbra''\ldots).
\end{itemize}
Note that by definition
\beq (\f|\xi>=\bbar{<\xi|\f)}. \eeq
The operators are elements of $L^{\times}(\F;\F')$, a space
containing also the spaces $L(\F)$, $L(\Hc)$, $L(\F')$, $L(\F;\Hc)$
and $L^{\times}(\Hc;\F')$ (after identifying $\Hc$ and $\F$ as subspaces of 
$\F'$ and restricting mappings from $\Hc$ and $\F'$ to $\F$).

\subsection{Definitions}

The map $f\mapsto f'$ defined by
\beq  <f'|\f):=(f,\f),\;\forall\f\in\F  \eeq
is the natural (antilinear) embedding of $\Hc$ as a (sequentially)
dense subspace of $\F'$. (The prime in $f'$ is usually dropped.)

\noindent $A^{*}$ denotes the Hilbert-space-adjoint of an operator $A$
in $\Hc$.

\noindent $A\dg\in L^{\times}(\F;\F')$, ``the adjoint'' of
$A\in L^{\times}(\F;\F')$, is defined by
\beq  <A\dg\psi|\f):=(\psi|A\f>,\;\forall\f,\psi\in\F.  \eeq
If $A(\F)\subset\Hc$ (so that $A^{*}$ is uniquely defined), then for
each $\f\in\F$
\beq  A^{*}\f\mbox{ is defined}\Longleftrightarrow A\dg\f\in\Hc  \eeq
and in this case $A^{*}\f=A\dg\f$, so $A\dg$ is the extension to all of
$\F'$ of the restriction of $A^*$ to $\F$.

\noindent $B'\in L(\F')$, ``the dual'' of $B\in L(\F)$, is defined by
\beq  <B'\xi|\f):=<\xi|B\f),\;\forall\f\in\F,\xi\in\F'  \eeq
and satisfies, for each $\psi\in\F$
\beq  B'\psi'=B\dg\psi\mbox{ ($=(B^{*}\psi)'$, when defined)}  \eeq
so $A'$ extends $A\dg$ (and thus also $A^*$) from $\F$ to $\F'$.

\noindent The following definitions are made:
(for $B\in L(\F),\f,\psi\in\F,\xi\in\F'$)
\begin{eqnarray}
  <\xi|B|\f)  & := & <\xi|B\f)\;(=<B'\xi|\f)\;)  \\
  (\psi|B|\f) & := & <\psi'|B|\f)=(\psi,B\f)\label{sym}
\end{eqnarray}
and when $B^{*}\psi$ is defined, also
\beq\label{conA-def}
  (\psi|B|\xi>\;:=\;\bbar{<\xi|B^{*}|\psi)}=(B^{*}\psi|\xi>.  \eeq

\noindent The (generalized) basis of bras for $\F$ suggested naturally
by this construction is:
\beq  \{<x|\;|x\in\Om\}  \eeq
where (for $\mu$-almost all $x\in\Om$) $<x|$ is defined by
\beq\label{xbra}  <x|\f):=\f(x),\;\forall\f\in\F.  \eeq
This is a ``complete, orthonormal'' \cite{Parseval}
system of eigenvectors of $A$ in the sense that the following
relations are satisfied:
\begin{enumerate}
  \item The Parseval equality:
    \begin{equation}\label{pars}
      (\psi,\varphi)=\int_{\Omega}d\mu(x)(\psi|x><x|\varphi),\;
        \forall\varphi,\psi\in\Phi.
    \end{equation}
  \item The eigenvalue equation:
    \begin{equation}
      <x|A_{i}\varphi)=x_{i}<x|\varphi),\;
        \forall\varphi\in\Phi,x\in\Omega
    \end{equation}
    or more briefly
    \begin{equation}
      A_{i}'\xi_{x}=x_{i}\xi_{x}
    \end{equation}
\end{enumerate}
Generalizing (\ref{xbra}), we denote (for $\xi\in\F'$)
\beq
  <\xi|x>\equiv\xi(x),\hspace{10mm}
  <x|\xi>\equiv\bar{\xi}(x),
\eeq
the generalized ``values'' of the generalized functions $\xi$ and
$\bar{\xi}$. ($\bar{\xi}$ is defined by \linebreak
$<\bar{\xi}|\f):=\bbar{<\xi|\bar{\f})}$.)

\noindent Finally, the matrix elements $<x|A|y>$ of an operator
$A\in L^{\times}(\F;\F')$ are defined to be the generalized ``values''
of the kernel $<A>\in\D'(\Om\times\Om)$ satisfying
\beq\
  (\psi|A|\f)=\int_{\Om}d\mu(x)\int_{\Om}d\mu(y)(\psi|x><x|A|y><y|\f).
\eeq
(Such a kernel does exist, by ``the kernel theorem'' of Schwartz,
which states \cite[p. 531]{Treves} that there is a natural isomorphism
between $L^{\times}(\Phi;\Phi')$ and ${\cal D}'(\Omega\times\Omega)$.)

\subsection{The Rules}

All the above definitions obey the following three rules:
\begin{enumerate}
  \item Whenever two of the expressions $(f|g>$, $<f|g)$ and
    $(f,g)$ are defined \linebreak
    (\ie when $f,g\in\Hc$ and at least one of them is in $\F$), they
    are equal:
    \beq
      (f|g>=<f|g)=(f,g)=\int_{\Om}d\mu(x)\bbar{f(x)}g(x).
    \eeq
    The same is true for $(f|B|g>$, $<f|B|g)$ and $(f,Bg)$.
  \item The conjugate of a bracket product (\eg $<\;|\;)$
    or $(\;|\;|\;)$ ) is the product of the conjugates in reverse order
    (if this last product is defined), where under conjugation \ldots:
    \beq\begin{array}{lcc}
      \f\in\F & : & (\f|\longleftrightarrow|\f)  \\
      \xi\in\F' & : & <\xi|\longleftrightarrow|\xi>  \\
      A\in L^{\times}(\F;\F') & : & A\longleftrightarrow A\dg
    \end{array}\eeq
  \item In any well defined expression of the form
    $(\psi|A_{1}\ldots A_{k}|\f)$ or $<x|A_{1}\ldots A_{k}|\f)$ or
    $<x|A_{1}\ldots A_{k}|y>$, the expression
    \beq  I:=\int_{\Om}d\mu(x)|x><x|  \eeq
    can be formally inserted between any two factors to obtain a
    decomposition in terms of components and matrix elements
    (note that the Parseval equality (\ref{pars}) is the simplest example 
    of this rule). Mathematically, $I$ is the embedding of $\F$ into 
    $\bbar{\F}'$ and of $\bbar{\F}$ into $\F'$.
\end{enumerate}
This is very close to the original rules introduced by Dirac, but there are
some complications:
\begin{itemize}
  \item The conjugate of a bra $<\xi|$ is $|\xi>$, which is also
        a bra and in general not a ket (since not every bra has a
        corresponding ket) so the conjugate of a ket $|\varphi)$
        should be seen as a ket $(\varphi|$ (this is why the kets and
        the bras are denoted by different symbols).
  \item The conjugate of an operator
        $A\in L^{\times}(\Phi;\Phi')$ is the adjoint $A^{\dagger}$.
        This works fine between kets but between a bra and a ket
        (when $A\in L(\phi)$), for the conjugate expression to be
        defined, the Hilbert-space-adjoint $A^{*}$ must be defined
        (see eq. (\ref{conA-def}) ).
  \item In the original Dirac formalism $I$ is the identity operator and not an
        embedding, therefore this formalism is not recovered here fully; not 
        every expression allowed by the formalism is well defined.
\end{itemize}

\newsection{The Group $\Pc(r,s)$}

${\cal O}(r,s)$, called ``the pseudo-orthogonal group of signature
$(r,s)$'', is the group of all linear transformations in
${\bf R}^{r+s}$
\begin{equation} x\mapsto x'=\Lm x \end{equation}
(where $\Lm$ is a real $(r+s)\times(r+s)$ matrix) that conserves the
quadratic form $x\cdot x$, where
\begin{equation}
  x\cdot y\equiv x_{\mu}y^{\mu}\equiv\eta_{\mu\nu}x^{\mu}y^{\nu} \hsc
  \eta=\mbox{diag}(\overbrace{1,\ldots,1}^{s},
                   \overbrace{-1,\ldots,-1}^{r}).
\end{equation}
$\Pc(s,r)$, called ``the inhomogeneous pseudo-orthogonal group'', is
the group of all affine transformations
\begin{equation}\label{Pmap}
  x\mapsto x'=\Lm x+a\mbox{, where }
    \Lm\in{\cal O}(r,s),\;a\in{\bf R}^{r+s}.
\end{equation}
The composition low in $\Pc(r,s)$ is (according to eq. (\ref{Pmap}) )
\begin{equation}
  (\Lm_{2},a_{2})(\Lm_{1},a_{1})=(\Lm_{2}\Lm_{1},\Lm_{2}a_{1}+a_{2})
\end{equation}
and the inverse is
\begin{equation}
  (\Lm,a)^{-1}=(\Lm^{-1},-\Lm^{-1}a)
\end{equation}
thus $\Pc(r,s)$ is the semidirect product of ${\cal T}_{r+s}$, the
translation group in ${\bf R}^{r+s}$, and of ${\cal O}(r,s)$.

${\cal O}\so(r,s)$ and $\Pc\so(r,s)$ denote the identity component
(the largest connected subgroup) of ${\cal O}(r,s)$ and
$\Pc(r,s)$ respectively and $\bbar{\cal O}\so(r,s)$ and
$\bbar{\Pc}\so(r,s)$ denote their universal covering groups.
To symplify notation, $\Pc\so(r,s)$, $\bbar{\Pc}\so(r,s)$,
${\cal O}\so(r,s)$, $\bbar{\cal O}\so(r,s)$ and ${\cal T}_{r+s}$
will be abbreviated by $\Pc\so,\;\Pc,\;\Lc\so,\;\Lc$ and $\cal T$
respectively.

There is a homomorphism from $\Lc$ onto $\Lc\so$. Its kernel is
\cite{Barg-ray} ${\cal N}(r)\otimes{\cal N}(s)$ where ${\cal N}(r)$ is
a cyclic group:
\beq
  {\cal N}(r)=\left\{
  \begin{array}{lc}
    {\bf Z}_{1}\;\; & r=1      \\
    {\bf Z}         & r=2      \\
    {\bf Z}_{2}     & r\geq3.
  \end{array}\right.
\eeq

$\cal T$ is connected and simply connected, therefore
\begin{itemize}
  \item $\Pc\so$ is a semidirect product of $\cal T$ and $\Lc\so$.
  \item $\Pc$ is a semidirect product of $\cal T$ and $\Lc$.
  \item The homomorphism from $\Lc$ onto $\Lc\so$ extends naturally
    to a homomorphism from $\Pc$ onto $\Pc\so$, with the same kernel,
    therefore $\Pc\so$ is the quotient group
    \beq  \Pc\so=\Pc/({\cal N}(r)\otimes{\cal N}(s))  \eeq
\end{itemize}
The homomorphism from $\Pc$ onto $\Pc\so$ identifies any (projective
or true) representation of $\Pc\so$ as a (projective or true)
representation of $\Pc$. Such a homomorphism, from $\Pc$ to $\Pc'\so$
(also with an abelian discrete kernel), exists for any connected
group $\Pc'\so$ which is locally isomorphic to $\Pc\so$.

\subsection{Induced Representations of $\Pc$}

$\cal T$ is an Abelian group (isomorphic to ${\bf R}^{r+s}$) and thus
its dual $\T$ -- the set of non-equivalent irreducible continuous
unitary representations of $\cal T$ -- consists of characters:
\begin{equation}
  \T=\{\chi_{p}:a\mapsto e^{ip\cdot a},\;\forall a\in{\cal T}
       |p\in{\bf R}^{r+s}\}.
\end{equation}
$\T$ is isomorphic to ${\bf R}^{r+s}$ and will be called ``the
momentum space'' (following the physicists' terminology in the 4-D
case). $\Pc$ acts naturally on $\T$:
\begin{equation} (\Lm,a):p\mapsto\Lm\so p \end{equation}
where $\Lm\so\in\Lc\so$ is the image of $\Lm\in\Lc$
(notice that from now on $\Lm$ denotes an element of $\Lc$
and not of $\Lc\so$, although it corresponds to an element of $\Lc\so$).
$\cal T$ acts trivially, so we can say that only $\Lc$ acts. This
action will be denoted simply by $\Lm p$. It decomposes $\T$ into
orbits. These are classified according to the
``stability group'' $\Lc(p)=\{\Lm\in\Lc|\Lm p=p\}$ of their elements
(called also ``isotropy group'' or ``little group''). To classify the
orbits, we denote \cite{p-note}:
\begin{description}
  \item[$p_{+}$] the vector of the first $s$ components of $p$,
  \item[$p_{-}$] the vector of the last  $r$ components of $p$.
\end{description}
If $r,s>1$ the orbits are characterized by
$p_{\mu}p^{\mu}=p_{+}^{2}-p_{-}^{2}$ ($p_{\mu}p^{\mu}=0$ splits to two
orbits: $p=0$ and $p\neq0$). If $s=1$, some of the orbits split to
two: $p_{+}>0$ and $p_{+}<0$ and similarly for $r=1$. Table B.1 lists
the orbits, a representative $\po$ of each orbit and the little group
of $\po$.
\begin{table}
  \vspace{11.0cm}
  \caption{The orbits of $\Lc\so={\cal O}\so(r,s)$}
  \label{table}
  \vspace{10.0cm}
\end{table}

Given an orbit $\Tm$ (for simplicity, unless otherwise stated, the 
subscript $m$ is here an abstract symbol of an arbitrary type of orbit and 
not necessarily the time-like mass $m=\sqrt{p_{\mu}p^{\mu}}$), 
a representative $\psm\in\Tm$ and a unitary representation $L$
of $\Lc(\psm)$ in a (complex separable) Hilbert space $\Hc(L)$, $L$
induces a representation $U$ of $\Pc$ in $\Lsq_{\mm}(\Tm,\Hc(L))$,
where $\mm$ is the
non trivial $\Lc\so$-invariant Radon measure on $\Tm$ (which exists,
since both $\Lc\so$ and $\Lc\so(\psm)$ are unimodular, and is unique
up to a multiplicative constant). First, one chooses for each
$p\in\Tm$, an element $\Lmp$ of $\Lc$ that satisfies $\Lmp\psm=p$, and
then $U$ is defined by:
\begin{equation}
  [U(\Lm,a)f](p):=e^{ip\cdot a}L(\Dl(\Lm,p))f(\Lm\inv p)
\end{equation}
\[ \mbox{where }\Dl(\Lm,p):=\Lmp^{-1}\Lm\Lm_{\Lm^{-1}p}\in\Lc(\psm). \]
For convenience, we choose $\Lm_{\psm}=1$ to obtain
\begin{eqnarray}
  {}[U(\Lmp)f](p)=f(\psm),& &\forall p\in\Tm \\
  {}[U(\Dl)f](\psm)=L(\Dl)f(\psm)& &\forall\Dl\in\Lc(\psm).
\end{eqnarray}
$U$ is denoted by
\begin{equation}
  U=(\chi_{\psm}L)(\Pc(\psm))\uparrow\Pc
\end{equation}
indicating that it is a representation of all of $\Pc$ ``lifted'' from
a representation $\chi_{\psm}L$ of a subgroup $\Pc(\psm)$ of $\Pc$.
This is a unitary representation and if $L$ and the mapping
$p\mapsto\Lmp$ are continuous then so is $U$.

The set $\{\psm\}$ of representatives of the orbits (as chosen in
the table) is obviously a measurable set in $\T$ so, according to
Mackey's theorem \cite[p. 279]{Maurin}, every unitary representation of
$\Pc$ is equivalent to an induced representation and it is irreducible
iff the inducing representation (the one on the little group) is
irreducible. Thus an irreducible unitary representation on $\Pc$ is
characterized by an orbit and an irreducible unitary representation of
the little
group of this orbit. Combining this with the theory of decomposition
of a continuous unitary representation $U$ \cite[p. 162]{Maurin} and
with the assumption that the factors of $U$ are all of type I, one
obtains that the most general form of $U$ is as described in Section
2.2.

\subsection{Extending the Representation to Generators}

We denote:
\begin{description}
  \item[$\AP$]: the Lie algebra of $\Pc$ (generators of $\Pc$),
  \item[$\UP$]: the universal enveloping algebra of $\AP$ (polynomials
    in elements of $\AP$).
\end{description}
$\Pc$ is a separable smooth manifold so the spaces $\E(\Pc)$ and
$\E'(\Pc)$ (see Section A.1.2) are well defined.
$\E'(\Pc)$ is a *~algebra with respect to
\begin{description}
  \item[multiplication] (a ``convolution''):
    for each $T_{1},T_{2}\in\E'(\Pc)$, $T_{1}*T_{2}$ is defined by
    \begin{equation}
      \int_{\Pc}(T_{1}*T_{2})(x)f(x)dx:=\int_{\Pc\times\Pc}
        T_{1}(x)T_{2}(y)f(xy)dxdy,\;\forall f\in\E(\Pc)
    \end{equation}
\end{description}
and
\begin{description}
  \item[involution] (``conjugation''): The conjugate $T\dg$ of
    $T\in\E'(\Pc)$ is defined by
    \begin{equation}
      \int T\dg(x)f(x)dx:=\bbar{\int T(x)f\dg(x)dx},\;
        \forall f\in\E(\Pc)
    \end{equation}
    where $f\dg(x):=\bbar{f(x^{-1})}$.
\end{description}
Now $\AP$ is, by definition, the tangent space to $\Pc$ at the identity
$e$. Thus, in a coordinate neighborhood $(x_{1},\ldots,x_{n})$ of the
identity $x=0$, $A\in\AP$ is expressed by:
\begin{equation}
  Af=a_{i}\derp{x_{i}}f(x)|_{x=0}=-\int(a_{i}\derp{x_{i}}\dl(x))f(x)dx
    ,\;\forall f\in\E(\Pc)
\end{equation}
so $A$ acts in  $\E(\Pc)$ as a distribution: a linear combination of
derivatives of $\dl_{e}$ (the $\dl$-function with support in the
identity $e$ of $\Pc$: $\dl_{e}f=f(e)$ ). This identifies $\UP$
naturally with a sub-*-algebra of $\E'(\Pc)$:
\begin{equation}
  \UP\simeq\E'_{e}(\Pc):=\{T\in\E'(\Pc)|\mbox{supp}(T)\subset\{e\}\}
\end{equation}
which is spanned by derivatives of $\dl_{e}$, while $\AP$ is the
subspace of first order derivatives. We denote by
$T_{A}\in\E'_{e}(\Pc)$ the distribution corresponding to $A\in\UP$
and we get:
\begin{eqnarray}
  T_{AB}=T_{A}*T_{B}&&\forall A,B\in\UP \\
  T_{A\dg}=T_{A}\dg &&\forall A\in\UP
\end{eqnarray}
where the involution of $A\in\AP$ is defined by $A\dg:=-A$ and
extended to $\UP$ by \linebreak$(AB)\dg:=B\dg A\dg$. This means that the map
$A\mapsto T_{A}$ is a *~isomorphism from $\UP$ onto $\E'_{e}(\Pc)$.

$\Pc$ is also embedded naturally in $\E(\Pc)$ by the map
$x\mapsto T_{x}:=\dl_{x}$ in the following sense:
\begin{enumerate}
  \item $T_{xy}=T_{x}*T_{y},\;\forall x,y\in\Pc$ so $x\mapsto T_{x}$
    is an isomorphism,
  \item if $e^{tA}$ is a one-parameter subgroup of $\Pc$, generated by
    $A\in\AP$ then
    \begin{equation}
      T_{A}f=\der{t}f(e^{tA})|_{t=0}=
        [\int\der{t}\dl_{e^{tA}}(x)f(x)dx]|_{t=0}
    \end{equation}
    and this means that
    \begin{equation}
      \der{t}T_{e^{tA}}|_{t=0}=T_{A}=T_{\der{t}(e^{tA})|_{t=0}}.
    \end{equation}
\end{enumerate}

Now let $L$ be an irreducible unitary representation of $\Lc(\psm)$ in
$\Hc(L)$, inducing a representation $U$ on
$\Hc=\Lsq_{\mm}(\Tm;\Hc(L))$, as defined in  Section B.1, and suppose:
\begin{enumerate}
 \item $(\Psi,\Hc(L),\Psi')$ is a Gel'fand triple: $\Psi$ is a
   complete nuclear space, embedded in $\Hc(L)$ densely and continuously,
 \item the restriction of $L$ to $\Psi$ is a smooth representation of
   $\Lc(\psm)$ by continuous operators in $\Psi$:
   \begin{enumerate}
     \item $\Psi$ is $L$-invariant,
     \item for each $\Dl\in\Lc(\psm)$, $L(\Dl)$ is a continuous operator
       in $\Psi$,
     \item for each $\psi\in\Psi$, the function
       $\Dl\mapsto L(\Dl)\psi$ is a smooth function from $\Lc(\psm)$
       to $\Hc(L)$ (an element of $\E(\Lc(\psm);\Hc(L))$,
   \end{enumerate}
 \item the map $p\mapsto\Lmp$ is smooth on $\Tm$ (this can always be
   satisfied).
\end{enumerate}
Defining
\begin{equation} \Phi:=\D(\Tm;\Psi), \end{equation}
we get:
\begin{itemize}
  \item $(\Phi,\Hc,\Phi')$ is a Gel'fand triple with all the
    properties in (1),
  \item the restriction of $U$ to $\Phi$ is a smooth representation
    of $\Pc$ by continuous operators in $\Phi$.
\end{itemize}
The smoothness of $U$ in $\Phi$ allows the definition:
\begin{equation}\label{UtTf}
  \Ut(T)\f:=\int_{\Pc}d(\Lm,a)T(\Lm,a)U(\Lm,a)\f,\;
    \forall T\in\E'(\Pc),\f\in\Phi.
\end{equation}
Observing that
\begin{equation} U(\Lm,a)=\Ut(\dl_{(\Lm,a)}), \end{equation}
we see that the natural extension of $U$ to $\UP$ is
\begin{equation} U(A):=\Ut(T_{A}) \end{equation}
and it can be shown \cite[p. 2285]{Antoine2} that $U$ has the
properties 3, 4 and 5, enumerated in Section 2.2.

\newpage
Captions

Table \ref{table}: The orbits of $\Lc\so={\cal O}\so(r,s)$


\begin{thebibliography}{99}
\bibitem{Iagolnitzer}
  D. Iagolnitzer, {\em The S Matrix}, North-Holland 1978.
\bibitem{Dirac}
  P. A. M. Dirac, {\em The Principles of quantum Mechanics},
    Clarendon Press, Oxford, England, 1930 (1st eddition), 1947
    (3rd eddition).

\bibitem{Roberts-Bohm}
  J. E. Roberts, J.\ Math.\ Phys.\ {\bf 7} (1966) 1097; 
  Commun.\ Math.\ Phys.\ {\bf 3} (1966) 98.

  A. B\"{o}hm, {\em The Rigged Hilbert Space in Quantum Physics}, 
  in {\em Boulder Lectures in Theoretical Physics}, A.O.Barut 
  (ed.), vol. 9A (1966)

\bibitem{Antoine1}
  J.-P. Antoine, J. Math.\ Phys.\ {\bf 10} (1969) 53.

\bibitem{Antoine2}
  J.-P. Antoine, J. Math.\ Phys.\ {\bf 10} (1969) 2276.

\bibitem{RHS}
  see also:
 
  S. J. L. van Eijndhoven and J. de Graaf, {\em A Mathematical Introduction 
  to Dirac's Formalism}, North Holand, Amsterdam 1986.

  A. B\"{o}hm and M. Gadella, 
  {\em Dirac Kets, Gamov Vectors and Gel'fand Triplets}, 
  Lecture Notes in Physics, Springer-Verlag 1989.

  This approach has become standard in Quantum Field Theory; see for instance

  N. N. Bogolubov, A. A. Logunov and I. T. Todorov,
  {\em Introduction to Axiomatic Quantum Field Theory}, 
  Benjamin, New York 1975.

\bibitem{Horwitz-Pelc}
  L.P. Horwitz and O. Pelc, {\em Generalization of the Coleman-Mandula
  Theorem to Higher Dimension}, RI-2-96, TAUP 2175-94 preprint.
\bibitem{Barut}
  A. O. Barut and R. R\c{a}czka, {\em Theory of Group Representations
    and Applications}, PWN, Warszawa 1977.
\bibitem{isomorph}
  All the properties of $\Pc$ used in this article (mainly its relation to 
  $\Pc\so$) are invariant under isomorphism, so we take $\Pc$ to be 
  the universal covering group of $\Pc\so$ (and not only isomorphic to it).
  
\bibitem{Wigner}
  E. P. Wigner, Ann.\ Math.\ {\bf 40} (1939) 149; {\em Group Theory,}
  Academic Press Inc., N.Y. 1959.
\bibitem{Bargmann}        
  V. Bargmann, J. Math.\ Phys.\ {\bf 5} (1964) 862.
\bibitem{Barg-ray}
  Bargmann showed (V. Bargmann, Ann.\ Math.\ {\bf 59} (1954) 1) that for
  $r+s\geq3$, all the ray representations of $\Pc$ are (equivalent to) 
  true representations, so in this case assumption 2.2 is not a restriction).
\bibitem{As23}
  Assumption 2.3 means that $\Uone$ is expressible in terms of
  irreducible representations and since these are identified
  with particle types (see the discussion in the introduction of
  \cite{Horwitz-Pelc}), this requirement is actually part of the
  physical interpretation.
\bibitem{Nel-Sti}
  It can be shown \cite[p. 328]{Barut} that the
  Nelson-Stinespring criterion is satisfied by:
  \begin{description}
    \item[(a)] each element of $\AP$,
    \item[(b)] each central element of $\UP$,
    \item[(c)] each element of $\cal U_{K}$, where $\cal K$ is a
      compact and/or abelian subgroup of $\Pc$ (this includes (a) ),
  \end{description}
  so result R4 assures the existence of many essentially self
  adjoint operators among the representatives of $\UP$.
\bibitem{Maurin}
  K. Maurin, {\em General Eigenfunction Expansions and Unitary
    Representations of Topological Groups}, PWN, Warszawa 1968.
\bibitem{eqHn}
  The bar denotes the Hilbert space completion, which in this case means 
  the closure in $\Lsq_{\mu\n}(\Om\n)$, since it is a complete space. 
  The second equality in (\ref{Hn}) expresses the fact that 
  $\bigotimes_{1}\n\Lsq_{\mu}(\Om)$ is dense in $\Lsq_{\mu\n}(\Om\n)$.
\bibitem{Treves}
  F. Treves, {\em Topological Vector Spaces, Distributions and
    Kernels}, Academic Press Inc.\ New York, 1967.
\bibitem{Ires}
  Eq. (\ref{Par-n}) shows this for $\bigotimes_{1}\n\D(\Om)$,
  but this is a dense subspace of $\D(\Om\n)$ so the continuity of the
  bras extends (\ref{Par-n}) to all of $\D(\Om\n)$. In any case,
  because of the density of $\bigotimes_{1}\n\D(\Om)$ in $\D(\Om\n)$,
  usually only elements of $\bigotimes_{1}\n\D(\Om)$ are considered.
\bibitem{Stone}
  According to Stone's theorem (see M. Reed and B. Simon, {\em Methods of 
  Modern Mathematical Physics}, volume I: {\em Functional Analysis}, 
  Academic Press, N.Y. 1980, p. 265), if the
  function $t\mapsto(f,\Uone(g(t))h)$ is measurable for each
  $f,h\in\Hone$ (and particularly if $t\mapsto g(t)$ is continuous)
  then there exists a self adjoint operator $\Ag^{0}$ in $\Hone$ that
  satisfies $\Uone(g(t))=e^{it\Ag^{0}}$, which implies that
  \beq  \der{t}\Uone(g(t))h|_{t=0}=i\Ag^{0}h  \eeq
  (meaning that if for some $h\in\Hone$, one of the sides exists then
  both sides exist and they are equal). Comparing this to the
  definition of $\Ag$, one sees that for each $\f\in\Fone$:
  \[  \Ag^{0}\f\mbox{ is defined }\Longleftrightarrow\Ag\f\in\Hone  \]
  and in this case $\Ag^{0}\f=\Ag\f$ (we obtained such relation between
  $A^{*}$ and $A\dg$ in appendix A.4).
\bibitem{Gel'fand}
  I. M. Gel'fand, G. E. Shilov, and N. Ya. Vilenkin, {\em
    Obobshchennye Funktsii i Deistviya Nad Nimi}, Gosudarstvennoe
    Izdatel'stvo Fiziko-Matematicheskoi Literatury, Moskow, 1958-1960
    Vols.\ I-V\@. English translation: {\em Generalized Functions},
    Academic Press Inc., New York, 1964.
\bibitem{Jauch}
  J. M. Jauch and J.-P. Marchand (Helv.\ Phys.\ Act.\ {\bf 39} (1966) 325)
  seem to be the first to use the notation
  ``$<\cdot|\cdot)$'' but this was only for $\f(x)=<x|\f)$ and they did
  not consider $<x|$ as a distinct entity.
\bibitem{Reed}
  M. Reed and B. Simon {\em Methods of Modern Mathematical Physics},
    volume I: {\em Functional Analysis}, Academic Press, N.Y. 1980, p. 147.
\bibitem{Schwartz}
  L. Schwartz, {\em Th\'{e}orie des distributions}, Hermann \& Cie.,
    Paris, 1957-1959, Vols.\ I, II.
\bibitem{CSCO}
  The notion of a ``complete set of commuting observables'' may be formulated
  also in a pure algebraic manner, see:

  J.-P. Antoine, G. Epifanio and C. Trapani, 
  Helv.\ Phys.\ Act.\ {\bf 56} (1983) 1175.
\bibitem{Parseval}
  This is only in the sense of the Parseval equality, which
  is, in a Hilbert space, characteristic to a complete orthonormal
  system. This is not completeness in the usual sense since it is
  only in $\Phi$, while the $<x|$'s are not, in general, in $\Phi$.
  Orthogonality is not even defined since there is no inner
  product in $\Phi'$.
\bibitem{p-note}
  When $s=1$ \ie $\Lc\so={\cal O}_{0}(r,1)$, the standard notation for the 
  components of the momenta is:
  \[ p=(E,\pv\,) \hsc E=p^{0}(=p_{+}) \hsc
     \pv=(p^{1},\ldots,p^{r})(=p_{-}).  \]

\end{thebibliography}
\end{document}